# Freely Suspended Nematic and Smectic Films and Free-Standing Smectic Filaments in the Ferroelectric Nematic Realm

Keith G. Hedlund, Vikina Martinez, Xi Chen, Cheol S. Park,

Joseph E. Maclennan, Matthew A. Glaser, and Noel A. Clark

*Department of Physics, University of Colorado, Boulder, Colorado, 80309, USA*

Abstract

We show that stable, freely suspended liquid crystal films can be made from the ferroelectric nematic ($N_F$) phase and from the recently discovered polar, lamellar $SmZ_A$ and $SmA_F$ phases. The $N_F$ films display two-dimensional, smectic-like parabolic focal conic textures comprising director/polarization bend that are a manifestation of the electrostatic suppression of director splay in the film plane. In the $SmZ_A$ and $SmA_F$ phases, the smectic layers orient preferentially normal to the film surfaces, a condition never found in typical thermotropic or lyotropic lamellar LC phases, with the $SmZ_A$ films exhibiting focal-conic fan textures mimicking the appearance of typical smectics in glass cells when the layers are oriented normal to the plates, and the $SmA_F$ films showing a texture of plaquettes of uniform in-plane orientation where both bend and splay are suppressed, separated by grain boundaries. The $SmA_F$ phase can also be drawn into thin filaments, in which X-ray scattering reveals that the smectic layer planes are normal to the filament axis. Remarkably, the filaments are mechanically stable even if they break, forming free-standing, fluid filaments supported only at one end. The unique architectures of these films and filaments are stabilized by the electrostatic self-interaction of the liquid crystal polarization field, which enables the formation of confined, fluid structures that are fundamentally different from those of their counterparts made using previously known liquid crystal phases.

INTRODUCTION

The synthesis of new, highly polar, rod-shaped mesogens, RM734 by Mandle et al. [1] and DIO by Nishikawa et al. [2], and the resulting identification of the long sought-after ferroelectric nematic ($N_F$) phase [3,4,5], a spatially homogenous nematic liquid crystal (LC) with uniaxial polar ordering of its molecular dipoles and a large bulk polarization, has created a new area of soft matter research activity and interest. The identification of additional polar phases, such as the antiferroelectric lamellar smectic $Z_A$ ($SmZ_A$) and the uniaxial, polar ferroelectric smectic A ($SmA_F$), expanded and enriched the ferroelectric nematic realm. The $SmZ_A$, originally observed between the N and $N_F$ phases of DIO [2] and determined to be an antiferroelectric smectic phase with the molecular director and polarization parallel to the layers [6,7], has since been reported in several other materials [8]. The observation by the Stuttgart group [9] of $SmZ_A$-like textures in the Merck material AUUQU-2-N (2N) prompted the study of mixtures of 2N and DIO, where, in addition to the $SmZ_A$ phase, the $SmA_F$ phase [10] was found at lower temperature. The $SmA_F$ phase has also been observed in single-component analogs of DIO [11]. The structures of the $SmZ_A$, $N_F$, and $SmA_F$ phases are sketched in the Supplemental Information.

The discovery of the $N_F$ phase stimulated a number of theoretical and simulation studies of ferroelectric and related liquid crystal polar states, many of which are summarized in recent reviews on the subject [12,13]. The first atomistic simulation investigations of the molecular origins of polar order in $N_F$ materials identified a variety of characteristic molecular association motifs arising from electrostatic 'docking' interactions [3], a thermodynamic mechanism for polar order further explored in subsequent theoretical and simulation work [14,15,16,17]. The emerging consensus is that emergent polar order involves the subtle interplay of strong electrostatic association, excluded volume interactions, and molecular conformational behavior, but predictive modeling of phase behavior in this family of materials remains an unmet challenge. The ferroelectric nematic realm has proven to be extraordinary rich, and phenomenological theories of the phases, phase transitions, and properties of these novel polar materials are under active development [12,13, 18,19]. Ferroelectric nematic materials have potential use in a wide range of applications, including high speed electrooptic shutters and displays, non-linear optics, fast electronic electrooptic modulators, tunable lasers, electrostatic actuators, smart windows, shutters, quantum photonic devices, energy storage, optical imaging, and sensing [20,21,22,23,24,25,26,27,28].

Films and filaments are examples of LC preparations confined by interfacial tension at free surfaces and stabilized by LC nanostructuring. They have been found to be broadly useful in studying the effects of reduced dimensionality on liquid crystal structure and phase behavior. Single and few-layer thick freely suspended films of smectic liquid crystals, which have been studied extensively since their discovery in the 1970's [29], are confined along a single spatial dimension, yielding a unique platform for investigating two-dimensional (2D) elastohydrodynamics, interfacial effects, and phase behavior [30]. The surfaces of smectic films are comprised of smectic layers, nanostructured interfaces that suppress pore formation and film rupture. Nematic LC films can be also drawn but are metastable because of their susceptibility to rupture by capillary thinning.

Freely suspended filaments can be drawn from a variety of liquid crystal systems, including discotic phases where the molecules are stacked in fluid columns aligned along the filament axis [31,32], and bent-core smectic [33,34,35,36] or twist-bend phases [37], in which lamellae are arranged in concentric cylinders about the filament axis. Freely suspended filaments of nematics, and of smectics of rod-shaped molecules, tend to be very short-lived, thinning and breaking because of the Rayleigh-Plateau instability [38]. However, it has recently been found that this instability can be suppressed in ferroelectric nematic filaments by applying an electric field along the filament axis. For example, in the course of exploring the use of ferroelectric nematics as thick, liquid bridges in electrostatic actuators [39], Nishimura et al. observed the transient formation of freely suspended $N_F$ filaments between planar electrodes. More recently, Máthé et al. and Jarosik et al. independently reported that the lifetime of filaments of ferroelectric nematics up to several mm in length could be extended by the appropriate application of electric fields [40,41], enabling their detailed study.

Confirmation of the smectic nature of the SmZ$_A$ and SmA$_F$ phases by X-ray diffraction [7,10]

raised the question of whether such lamellar phases could be drawn into films or filaments that, like soap films, are "freely suspended" by their own surface tension. In this paper, we show that stable, freely suspended films can indeed be created not only in these polar smectic phases but also, remarkably, in the $N_F$ phase. In addition, we show that thin, polar filaments can be drawn in the $SmA_F$ phase, and that such filaments can be "free standing", *i.e.*, mechanically stable in spite of their own surface tension. These unique characteristics arise from the dominant electrostatic self-interaction of the polarization field in these highly polar materials, which leads to film and filament structures that are fundamentally different from those of their counterparts made using previously known LC phases.

*RESULTS*

In these experiments, we studied a 50:50 weight-percent (wt%) mixture of the mesogens DIO and AUUQU-2-N [42], which exhibits the bulk phase sequence Isotropic (144°C) N (83.7°C) $SmZ_A$ (66°C) $N_F$ (55°C) $SmA_F$ [10]. Freely suspended films of this mixture were drawn at temperatures in the $SmZ_A$ phase range across a circular hole in a glass cover slip and could subsequently be cooled into the $N_F$ and $SmA_F$ phases. These films were observed in transmission, with typical textures seen between crossed polarizer and analyzer shown in Fig. 1. Extensive optical microscopy of this mixture in planar glass cells confirms that in these phases the director ***n*** and polarization ***P*** are parallel to one another.

*Films – $SmZ_A$ phase* – At temperatures in the $SmZ_A$ phase, the film texture viewed in polarized transmitted light is reminiscent of the focal conic fan textures of smectics with bookshelf layering in sandwich cells with weak or no alignment  The birefringence color is locally uniform but gradually changes near the edge of the film where the thickness *d* increases. Assuming a typical $N_F$ birefringence, the color indicates this film is around 300 nm thick in the center. The presence of dark extinction brushes here (and, indeed, in $N_F$ and $SmA_F$ films) indicates that the director, and therefore the polarization, is locally uniform through the film and oriented parallel to the film surface. In the $SmZ_A$ phase, where the director is parallel to the smectic layers, this alignment could, in principle, be achieved with the layers either parallel or normal to the film plane. The smectic-like domain patterns would indicate that the SmZ$_A$ layers are, in fact, normal to the film plane, the bookshelf geometry also preferred in cells [6, 7].

*Films – N$_F$ phase* - The transition on cooling to the N$_F$ phase is first-order. The film texture in the N$_F$ phase is much smoother than in the SmZ$_A$, with both the in-plane director orientation and the birefringence color uniform over large areas, as seen in Fig. 1B. The presence of black extinction brushes indicates again that the director is parallel to the film plane and uniform through the thickness of the film. The texture is decorated with parabolic and hyperbolic defect lines of distinctly lower birefringence which may anneal away, leaving a smoothly varying 2D director field of the kind illustrated in Fig. 1C.

We now consider the origin of the preference for **n** being parallel to the film plane, noting that this orientation of the director is not generally found in nematic free films and has never been observed in freely suspended smectic films of calamitic or bent-core molecules, where the layers align parallel to the surfaces so that the director is consequently oriented normal (or nearly normal) to the surfaces. We can model the N$_F$ film as a uniformly polarized slab with a coupled director/polarization field **n**($\theta$, $\varphi$)/**P**($\theta$,$\varphi$) making an angle $\theta$ relative to the film surface and an orientation $\varphi$ in the *y,z*-plane of the film. Polarization self-interactions tend to suppress splay distortions of the polarization field **P**, so that the director/polarization field is essentially uniform through the thickness of the film [29,43,44,45]. Tilting of such a uniform polarization block from $\theta = 0$ would deposit space charge of magnitude $P\sin\theta$ and with opposite signs on the two film surfaces. This in turn would generate a uniform field through the thickness of the film, normal to the film plane, of magnitude $E = -(P/\varepsilon)\sin\theta \sim (10^9 \text{ V/m})\sin\theta$, the energetic cost of which effectively suppresses any deviation from planar orientation. The polarization self-interaction thus forces **P** to be both locally uniform through the film and parallel to the film plane, a phenomenon also observed for N$_F$ material confined in channels [18]. A similar condition applies in the antiferroelectric SmZ$_A$ phase, where a uniform polarization field is preferred within each smectic layer [6,7].

The texture of freely suspended films in the $N_F$ phase may thus be considered to be an image of the two-dimensional director/polarization field, $\varphi(y,z)$. The texture is controlled only by the LC elasticity and the self-interaction of the polarization field, which, due to polarization space charge, makes in-plane splay deformations much more energetically costly than bend. This is manifest in the films as a tiling of large areas within which *n*/***P*** is uniform (blue/green regions in Fig. 2A) or is tangent to families of circles, exhibiting pure bend, with the required complementary splay being confined to narrow defect lines (polarization-stabilized kinks, curved yellow/gray lines in Figs. 1B,2) [43,44,45]. The constraint allowing only bend of the director is the conjugate of the constraint permitting only director splay in 2D smectics [46], both of which lead to textures of 2D focal conics with characteristic defect lines in the form of parabolas or hyperbolas, as illustrated in Fig. 2. Similar free-boundary conditions are achievable in supported $N_F$ films bounded by air on top and a glycerin or untreated polymer surface on the bottom [47,48]. These preparations all yield textures very similar to that shown in Fig. 1B. In contrast to conventional smectic A and C films [30], the thickness of the $SmZ_A$, $N_F$, and $SmA_F$ films is not quantized in the sense of having a certain number of smectic layers in a given location, and the characteristic island and layer-step textures found in conventional smectic films are not observed here. Instead, as is evident from Fig. 1, the thickness of the films varies continuously across the width of the film, with the birefringence colors indicating that the films thicken significantly on approaching and entering the meniscus.

Striking behavior related to film thickness is observed in the $N_F$ phase upon cooling, where the films thin dramatically near the point defects of the 2D focal conic texture, as illustrated in Fig. 3A. These topological defects comprise a region of circular bend of the *n*/***P*** field where the director is tangent to families of concentric circles, making $2\pi$ circuits around a common center so they are strength +1 disclinations in $\varphi(y,z)$. The area around the required companion -1 defects breaks up into plaquettes of uniform orientation separated by line singularities [43]. As the films are cooled, local thinning around these defects leads to a mutual attraction, which causes the defects to collect spontaneously into distinct clusters. The focal conics become distorted in this process, with the parabolic defects eventually being stretched into nearly linear domain walls connecting different clusters of point defects. The film thickness $d$ in the defect clusters of Fig. 3 is only around 100 nm, substantially thinner than the roughly 800 nm-thick, blue-green region surrounding them. The thinning is correlated with the degree of in-plane curvature of the director field, with $d$ beginning to decrease strongly from the background film thickness on approaching the outer reaches of a defect where the circular director pattern gets established, then decreasing further on approaching the defect core, so that the films are thinnest near the defect cores, as seen in Fig. 3A. Further illustrations of defect clustering may be seen in the Supplemental Information.

These observations suggest the following model for film thinning and defect clustering: (*i*) the intrinsic film stability indicates a thickness-dependent balance of disjoining pressure contributions in a film tending respectively to thin or thicken it; (*ii*) the defects have associated excess energy, in the form of director curvature energy/film area, $U_d$, and defect core energy, $V_d$, both of which are proportional to film thickness. These energies are reduced if the film becomes thinner, resulting in an excess contribution to the disjoining pressure that acts like a spring connecting the film surfaces, pulling them together; (*iii*) this tendency extends to the areas between the defects, resulting in a mutual attraction analogous to the capillary attraction of nonwetting particles on water that depresses the surface [49]. (*iv*) this action causes the films to thin but not to the point of rupture, indicating that thinning is opposed by a repulsive pressure that increases with decreasing thickness. The origin of this pressure is not known but could originate from the film being charged relative to the surrounding bulk LC on the film holder. Such charge is expected because the meniscus at the film edge is essentially polar, varying in thickness and structure with increasing radius from the film edge to the bulk material, such that there is always some radial component of **P** of preferred sign, which will leave the film charged.

*Films – SmA$_F$ phase* – The global textures, film thickness, and birefringence do not change significantly as the film is cooled from the N$_F$ into the SmA$_F$ phase, but locally all regions of the film show a remarkable evolution from continuous reorientation of the director in the plane of the film to discontinuous reorientation, as seen by comparing Fig. 1C with Fig.1D, where a film having a smooth variation of orientation and thickness has broken up into irregular-sized blocks bounded by distinct linear boundaries that sharply demarcate different in-plane orientations. The film thickness may still change smoothly within each block but the orientation is uniform or nearly so within each block. This tendency toward uniform local orientation is a manifestation of the presence of smectic layers. Since the polarization is parallel to the film plane and the layers normal to it, the texture corresponds to a 2D arrangement of uniformly spaced layers spanning the thickness of the film. In discussing the focal conic texture of the $N_F$ phase, we argued that director/polarization splay is suppressed by space charge such that the areas remote from the line defects have only director bend. However, following the arguments of Friedel [50] and deGennes [51], in smectic phases with the director normal to the layers, such as the $SmA_F$, director bend must also be expelled, so that the domains tend to have uniform in-plane director orientation, with neither splay nor bend, and areas of the film where there is bend curvature are forced to break up into uniformly aligned blocks. Dramatic examples of this are seen in the areas with large bend curvature immediately surrounding the +1 defect cores in the $N_F$ phase (Fig. 3A), which end up as a tiling of uniformly oriented plaquettes separated by grain boundaries in the $SmA_F$ phase (Fig. 3B). In general, the number of grain boundaries that form around the core of a point defect depends on their energy / length, with a minimum of two plaquettes (and two grain boundaries) required to preserve the topology of the defect. In the films, it appears that the minimum energy arrangement of grain boundaries is that having eight grain boundaries, each providing an orientational jump of 45° between each of eight plaquettes.

*$SmA_F$ Filaments – X-ray diffraction* – Filaments of 2N/DIO in the $SmA_F$ phase were drawn at either 45°C or 35°C by suspending a drop of LC between the tips of a pair of coaxial metal needles and slowly separating the needles. Such filaments could made be up to several cm in length and tens of microns to 1 mm in diameter. Examples are shown in Figs. 4A and 5 and in the Supplemental Information. Experiments with a tunable birefringence compensator show that the ***n/P*** couple is along the filament axis, *z*, as was found in $N_F$ phase filaments [40,41]. This geometry has the polarization parallel to the LC/air interface, as expected on the basis of minimization of electrostatic energy as discussed above for the films, and the smectic $A_F$ layer normal oriented along the filament ($z$) axis.

X-ray diffraction was carried out in transmission using a laboratory-based SAXS/WAXS system with precision beam collimation and two-dimensional image acquisition. Both SAXS and WAXS images exhibited Bragg scattering from the SmA$_F$ layering in the filaments. The SAXS images (shown in Figs. 4A,B, respectively without and with an applied longitudinal electric field) give an arc of scattering that is resolution-limited in $q_z$ at $q_z = q_{AF} = 0.269$Å$^{-1}$ (layer spacing $d_{AF}$ = 23.4 Å), comparable to the $q_{AF} = 0.267$Å$^{-1}$ previously measured in the bulk SmA$_F$ phase of 2N/DIO [10]. The location of the scattering arcs confirms that the smectic layer normal is generally along the filament axis, but the azimuthal scans along these arcs shown in Fig. 4C indicate a rather broad mosaic distribution of layer normals, especially in the absence of an applied electric field (cyan curve). In an applied field (green curve), this distribution narrows and evolves to show several fairly sharp, individual peaks in $\theta$, indicating the presence of a few, well-ordered SmA$_F$ domains differing in orientation by a few degrees and filling the illuminated volume. This behavior is similar to domain patterns found in the SmA$_F$ in glass plate cells with alignment layers [10].

The WAXS scans (shown in Fig. 4D) revealed, in addition to the fundamental scattering from the SmA$_F$ layering at $q_z = q_{AF}$, a second-order SmA$_F$ layering peak at $q_z = 2q_{AF}$ and the fundamental scattering at $q_z = q_{CR}$ from a crystal phase with a layer spacing at $2\pi/q_{CR} \approx 41$Å. All of these scattering features have peak intensity with $q$ along the filament axis $z$, so both the SmA$_F$ and the crystal layers are oriented generally normal to the filament axis. At lower temperatures (T = 35°C), the crystal scattering grows in intensity and the SmA$_F$ signal weakens over the course of about five minutes. As observed in SAXS, the scattering features in the WAXS scans consist of a collection of sharp sub-peaks that are distributed in azimuth. The WAXS sub-peaks also appear to be distributed in $q_z$. However, this spreading is a geometrical artifact arising from the finite depth of the scattering volume along the x-ray beam as it passes through the diameter of the filament, giving an apparent $q_z$ that depends on position of the scattering domain along the beam. This effect is largest when the detector is close to the sample (as in WAXS), and when the sample is thick (as for filaments). The radial scan of the SAXS data in Fig. 4B shows that the x-ray scattering from the layer structure of the SmA$_F$ phase in the filaments is in fact single-peaked.

The observation that the layer normal is generally parallel to the filament surface implies that the layers are oriented normal to the surface. This is in contrast to conventional freely suspended smectic A and C films, where it is the director that is oriented preferentially normal (or nearly normal) to the film surface. In such films, the smectic layers form parallel to the film surfaces and the film thickness is quantized (an integral number of smectic layers thick). An analogous arrangement, with the layers forming parallel to the surface, is seen in bent-core liquid crystal filaments in the B7 phase. Such stacked layers are readily detected by XRD. In the case of the SmA$_F$ filaments, however, there is no evidence of such scattering, which would show up in the regions indicated by orange circles in Fig. 4D. In the SmA$_F$ phase, the dominant determinant of director orientation at the LC/air interface is the elimination of surface charge, making *P* and hence *n* orient parallel to the surface. In a filament with a cylindrical surface, this boundary condition stabilizes planar layers normal to the filament axis. We will show below that in a SmA$_F$ filament with a conical surface, this boundary condition instead stabilizes conical layers oriented normal to the surface.

*SmA$_F$ Filaments – Conical Layering and Free-Standing Fluid Stacks* – In both freely suspended liquid crystal films and filaments, the air/LC surface tension acts as the suspension and stabilization mechanism, minimizing the surface energy by forming cylindrical or planar interfaces. However, surface tension can also act as a potential destabilizing mechanism, driving the instabilities that cause local thinning of films [30] or fluid filaments [38] and leading films to rupture and filaments to break. While such events are generally catastrophic for known fluid and LC filaments, this is not the case for filaments in the SmA$_F$ phase, which maintain stable, free-standing structures even after they are broken on one end.

Such a filament is shown in Figs. 5A,B,E where a filament initially drawn between two needles has broken and is now supported only by the left-hand needle. As the filament was drawn, it was both stabilized as a cylinder and thinned by Rayleigh-Plateau capillary forces but, in this example, these forces ultimately drove a topology-altering event, breaking the filament and leading to the formation of a hemispherical tip at the unsupported end where the Laplace pressure then acts to pump the fluid back into the filament, toward the contacting bulk material at the other end. Remarkably, broken SmA$_F$ filaments successfully resist this Laplace pressure, maintaining their extended cylindrical structure and not retracting, but bending downward in response to the earth's gravitational field, as seen in Fig. 5E. Images of a filament transitioning from freely suspended to free-standing after the very thin filament tethering the right-hand end breaks are shown in Figs. 5 C→E. The structural stability of these free-standing filaments tethered at only one end can be understood by analogy with confined stacks of discs of uniform thickness representing the smectic layers, as illustrated in Fig. S9 using NECCO candy wafers and discussed further below.

Let us consider a filament with typical dimensions, radius $r = 0.5$ mm, and length $L \sim 10$ mm. The surface tension, $\sigma$, through the Laplace pressure from the quasi-spherical end cap, compresses the smectic layering such that each layer applies a constant force $2\pi r\sigma$ to its neighbor along the filament axis. This compression holds the smectic layer normal parallel to the filament axis. The layers withstand the compression with only a small fractional reduction in layer spacing: $-\delta d_{AF}/d_{AF} = 2\sigma/rB \sim 10^{-4}$, where we have taken $\sigma \sim 20$ erg/cm$^2$ and $B \sim 10^7$ erg/cm$^3$ (the compressibility of 8CB, a typical small-molecule smectic near room temperature) [52].

The gravitational deformation of a broken filament such as the one depicted in Fig 5E can be compared to beam-bending: the response of a beam with uniform mass/length clamped at one end and loaded along its length by its own weight. Downward beam deflection results from two effects [53]: (*i*) the moment of torque transmitted along the beam, which deforms each vertical slice of the beam (each smectic layer) to be narrower at its bottom than at the top, a deformation opposed in the case of a smectic by the layer compressibility $B = E$, the effective Young's modulus; and (*ii*) shear strain $S_{xz}$, in which each vertical slice of the beam is pushed downward (along gravity) so that it slides on the neighboring material (the next smectic layer) on the side closer to the clamped end, a deformation for a solid beam that is opposed by its shear modulus $G$. For typical isotropic, solid beams with a length-to-width ratio $L/r \sim 10$, the shear contribution is negligible, and the downward beam deflection at the end is $\delta \approx (\rho g A L^4/8EI) \sim 0.2$ mm, where we have taken $\rho = 1$ g/cm$^3$, $g = 980$ cm/sec$^2$, area $A = \pi r^2$, and $I = (2r)^4/12$ [53].

A bulk smectic is not solid but has a bulk shear modulus $G = 0$, with each smectic layer able to slide like a fluid sheet relative to its neighbors, a key degree of freedom of fluid smectics. In a smectic A$_F$ filament, however, such motion is prevented by the shear rigidity provided entirely by the energetics of the boundary structure at the LC/air interface, which is what keeps the layers normal to the filament axis in the first place. The surface tension $\sigma$ puts the surface under extensive force along the cylinder axis, tending to keep the surface parallel to this axis. Additionally, the polarization is electrostatically constrained to be parallel to this surface, so that the fluid smectic layers orient perpendicular to the filament surfaces and terminate there. The surface tension and the electrostatics thus combine to maintain an interface that is locally smooth and locally normal to the layering plane. Shear strain $S_{xz}$ would tilt $P$ away from the filament axis (Fig. S9A). Tilt through a small angle $\theta \equiv S_{xz}$ would increase the electrostatic energy per unit filament volume by $U_{tilt} = \frac{1}{2}(P^2/4\varepsilon)S_{xz}^2$ [54], giving an effective shear modulus $G = (P^2/4\varepsilon) \sim 10^9$ ergs/cm$^2$, which is sufficiently large that shear is not a factor in filament deformation by gravity as described under criterion (*ii*) above.

If the smectic layers at the filament surface are constrained to be normal to the surface, the local filament shape in Fig. 5 can only be cylindrical or conical, where, in the conical case, shown in Figs. C and G, the layering too is conical, with the layers having the same inclination $\psi$ as the surface, as indicated in Fig. 5F. When the filament is tethered, surface tension along the filament minimizes deviations from cylindricality, and changes in diameter along its length are effected by the alternation of conical and cylindrical segments, manifested as a series of bright and dark reflection bands as seen in Fig. 5, where the cylindrical and conical segments are outlined respectively in red and green. The boundaries between cylindrical and conical surfaces are transition regions in which the entire local smectic layering changes from conical to planar. These transitions can be quite sharp and, judging from the images of the bands, take place in the same narrow range of $z$ across the diameter of the filament.

*SmA$_F$ Filaments – Optics of the Dark and Bright Bands* – The filament images shown here were obtained with a low-power, long working distance objective, which gives an imaging system with a relatively small aperture (3° cone angle) for collecting the reflected light. The filaments were imaged in depolarized light reflected principally from either the front or back surface of the filament. As is evident from the images in Fig. 5 and in the Supplemental Information, the dominant textural feature of the SmA$_F$ filaments is a pattern of alternating dark and bright bands. These bands are geometrical optical features that effectively image the surface orientation of the filament, as outlined below.

A ray diagram showing the geometrical optics of collecting this light, drawn for convenience in same plane as the filament image, is shown in Fig. 5F. (In reality, this plane of reflection ($x$,$z$) is, of course, normal to the image plane ($y$,$z$), with the incident light parallel to $x$.) Assuming the polarization of the incident light is parallel to the filament axis, the following observations can be made: (*i*) Fig. 5C shows a bright line running parallel to the filament along its center (the centerline). This feature is the back-reflection from the front surface of the filament, and is observed only from places where the filament front surface is normal to the incident light. Thus is narrowly confined to the filament center line since the filament curvature reflects other light away, and indicates that the topmost part of the filament is normal to the incident light in the x-z plane. (*ii*) Where the front surface of the filament is tilted, the filament curvature deflects the reflected light (magenta rays) out of the collection aperture so that reflections from the front surface are not collected and these regions appear homogeneously dark. Regions where the filament surface is normal to the incident light, on the other hand, reflect strongly into the collection optics, giving bright bands, for example along the cylindrical segments of an intact filament, as in Fig. 5C-F. When a filament breaks at one end, the filament typically tilts down slightly in gravity, behind the plane of the image, such that the back-reflected light is no longer collected. An example is shown in Fig. 5E, where the (invisible) fine tethering filament supporting the right-hand end of the main filament has broken and the center-line feature disappears. (*iii*) When the tether breaks and the thick remnant filament starts slowly bending down under gravity, the bands start moving continuously and smoothly along the filament toward the free end. When the filament is sufficiently tilted, the conical regions appear bright and the cylindrical regions dark, inverting the original contrast as can be seen by comparing Figs. 5C and 5E, this relative brightness determined by the direction of the back reflected light. These reflections appear fuzzy because of internal optical inhomogeneity of the filament. In principle, the bright bands can be translated at will along the filament simply by changing the angle of incidence of the incident light and the location of the collection aperture. The internal textures observed in free-standing filaments are otherwise similar to those of tethered filaments.

Finally, the bright bands are clearly not homogeneous: at higher resolution (Fig. 5G), they are seen to have a texture of fuzzy, bright lines running across the filament, apparently from SmA$_F$ layering defects, which are not unusual in samples many tens of microns thick. This striated texture is similar to that found in smectics confined between flat plates where temperature change has thinned the layers, leading to chevron formation and linear walls of zig-zag defects separating domains of opposite chevron sign [55]. An example of a similar texture in a cell in the smectic C phase is shown for comparison in Fig. 5G.

CONCLUSION

We have shown that highly polar liquid crystals in the SmZ$_A$, N$_F$ and SmA$_F$ phases form stable, freely suspended films. In contrast to films of traditional smectic materials, the layers are oriented preferentially normal to the film surfaces, the orientation preferred by electrostatic energy minimization at the surfaces. The SmZ$_A$ films exhibit focal conic fan textures mimicking the bulk textures of traditional smectics. N$_F$ films display parabolic focal conic textures that are a manifestation of the electrostatic suppression of director splay in two dimensions and show dramatic thinning around topological point defects. The SmA$_F$ films show a texture of uniform plaquettes separated by grain boundaries. SmA$_F$ material can also be drawn to form stable, linear, liquid crystal filaments between solid supports. The smectic layering is normal to the filament axis, again as preferred electrostatically, rather than adopting the concentric layer arrangement favored by surface tension previously observed in smectic filaments. Remarkably, SmA$_F$ filaments are stable even when broken, forming free-standing structures with only minor changes in their appearance.

METHODS

Freely suspended films were drawn across a 5 mm diameter, circular aperture in a microscope cover slip mounted in a temperature-controlled hot stage (a modified Instec Inc. Model HCS302). The films were drawn at 81°C, in the SmZ$_A$ phase, and viewed in transmission on an Olympus BX51 polarizing microscope. The film thickness, determined from the reflection spectrum [56], ranged from about 100 nm to 900 nm, the thickness depending on the phase (SmA$_F$ films tending to be the thickest) and the location in the film (all films tending to be thicker near their edges).

Freely suspended SmA$_F$ filaments were created by applying a small amount of the DIO/2N mixture in the nematic phase (at 130°C) to the tip of a horizontal metal needle mounted inside a temperature-controlled hot stage (Instec Inc. Model HCS302) held at either 45°C or 35°C. The needle tip, coated with material now in the SmA$_F$ phase, is touched to a second, colinear needle mounted within the hot stage and then slowly retracted, creating a thin filament up to several cm in length stretching between the two needles. Applying a longitudinal 10 V, 2 Hz triangle-wave voltage to the filament during the drawing process greatly increased the initial stability of the filament. The electric field could be removed once the filament had been fully drawn.

The filaments were viewed in polarized reflected light on an Olympus BX51 microscope. X-ray diffraction experiments were carried out using a Forvis SAXS/WAXS diffractometer with a 30 W Xenocs Genix 3D X-ray source (Cu anode, $\lambda$ = 1.54 Å) and a Dectris Eiger R 1M detector.

ACKNOWLEDGEMENTS

This work was supported by NASA Grants NAGNNX07AE48G and NNX-13AQ81G, and by the Soft Materials Research Center under NSF MRSEC Grants DMR 0820579 and DMR-1420736, and by NSF Condensed Matter Physics Grant DMR-2005170. K.G.H. was supported by a UROP research award from the University of Colorado Boulder. The authors are grateful to Frank Giesselmann at the University of Stuttgart and Matthias Bremer at Merck Electronics, Darmstadt for providing liquid crystal materials. X-ray experiments were performed in the Materials Research X-Ray Diffraction Facility at the University of Colorado Boulder (RRID: SCR 019304), with instrumentation supported by NSF MRSEC Grant DMR-1420736.

*AUTHOR CONTRIBUTIONS*

NC, JM, XC, MG, and CP conceptualized the project. KH, VM, and XC performed the experiments. KH, JM, NC, VM, XC, and MG analyzed and interpreted the data. KH, JM, and NC wrote the paper.

*Conflict of Interest Statement*

In accordance with University of Colorado policy and our ethical obligations as researchers, we are reporting that several authors have a financial interest in a company (Polaris Electro-Optics, Inc.) that may be affected by the research reported in this paper. We have disclosed those interests fully to the University of Colorado and have in place an approved plan for managing any potential conflicts arising from that involvement.

*Figures*

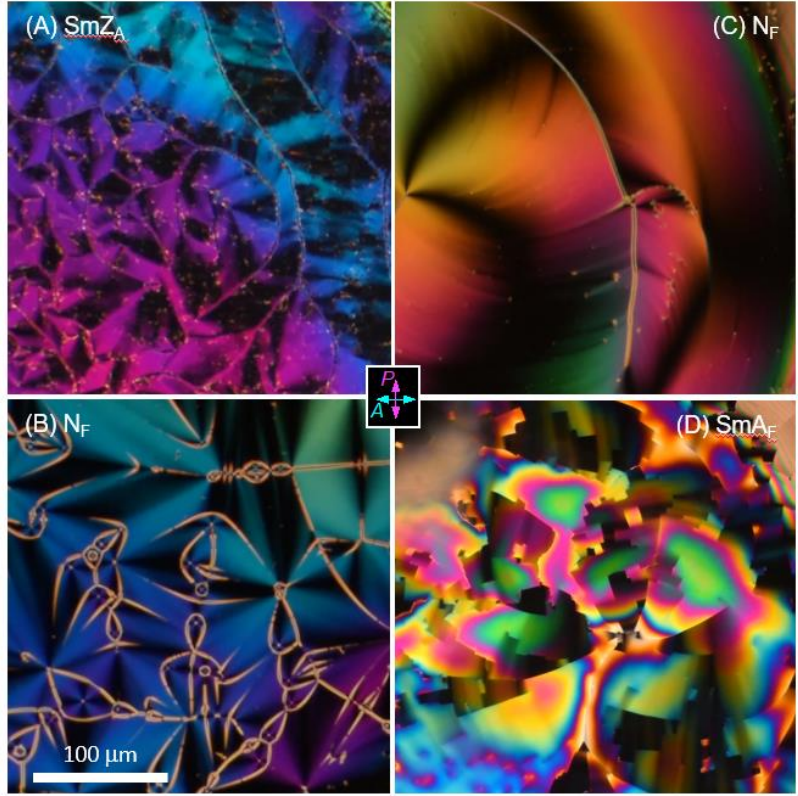

*Figure 1*. Typical polarized light microscope textures of a freely suspended 2N/DIO film in the (A) SmZ$_A$ (85°C), (B,C) N$_F$ (70°C, 60°C), and (D) SmA$_F$ (45°C) phases, viewed in transmission. In all phases, the director and polarization are parallel to the film plane and locally have the same orientation through the film, giving dark extinction brushes between crossed polarizers. The in-plane azimuthal orientation is determined by polarization self-interactions and elasticity. Variations in birefringence color are indicative of changes in film thickness. (A) The antiferroelectric, lamellar structure of the SmZ$_A$ phase leads to an irregular texture of small focal-conic fans. (B) In the N$_F$ phase, polarization splay is suppressed, leading to a bend-only texture. The central part of the image in (B) is shown in more detail in Fig. 2. (C) A large, circular N$_F$ bend domain. (D) The same area as in (C) after cooling to the SmA$_F$ phase. Here both bend and splay of the director are suppressed, leading to the formation of locally uniform, block-like smectic domains of different orientation, separated by sharp grain boundaries. Black domains are at extinction orientations.

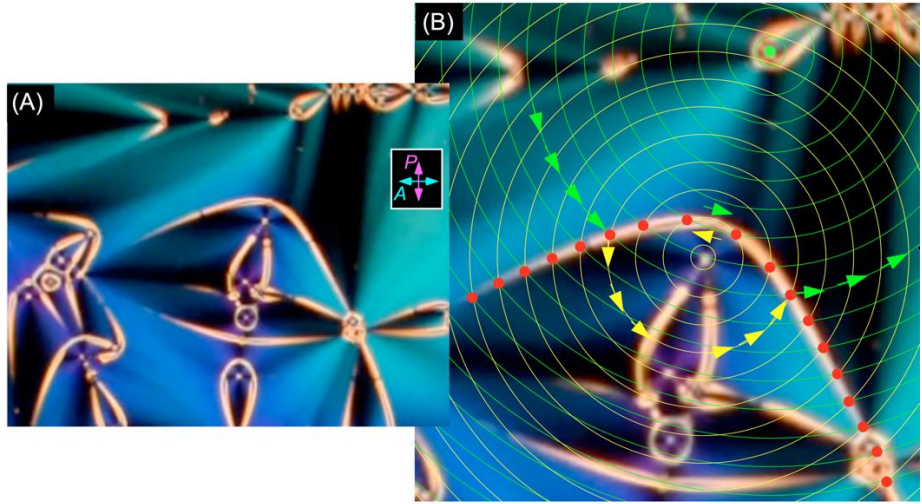

*Figure 2*. Magnified images of parabolic focal-conic line and point defects in the freely suspended 2N/DIO film in the $N_F$ phase shown in Fig. 1B. Two sets of circular contour lines, green and yellow, which the polarization/director field locally follows on opposite sides of the central parabola, are shown in (B). The parabolic focal-conic geometry guarantees that the director/polarization field, indicated schematically by colored arrows, is splay-free.

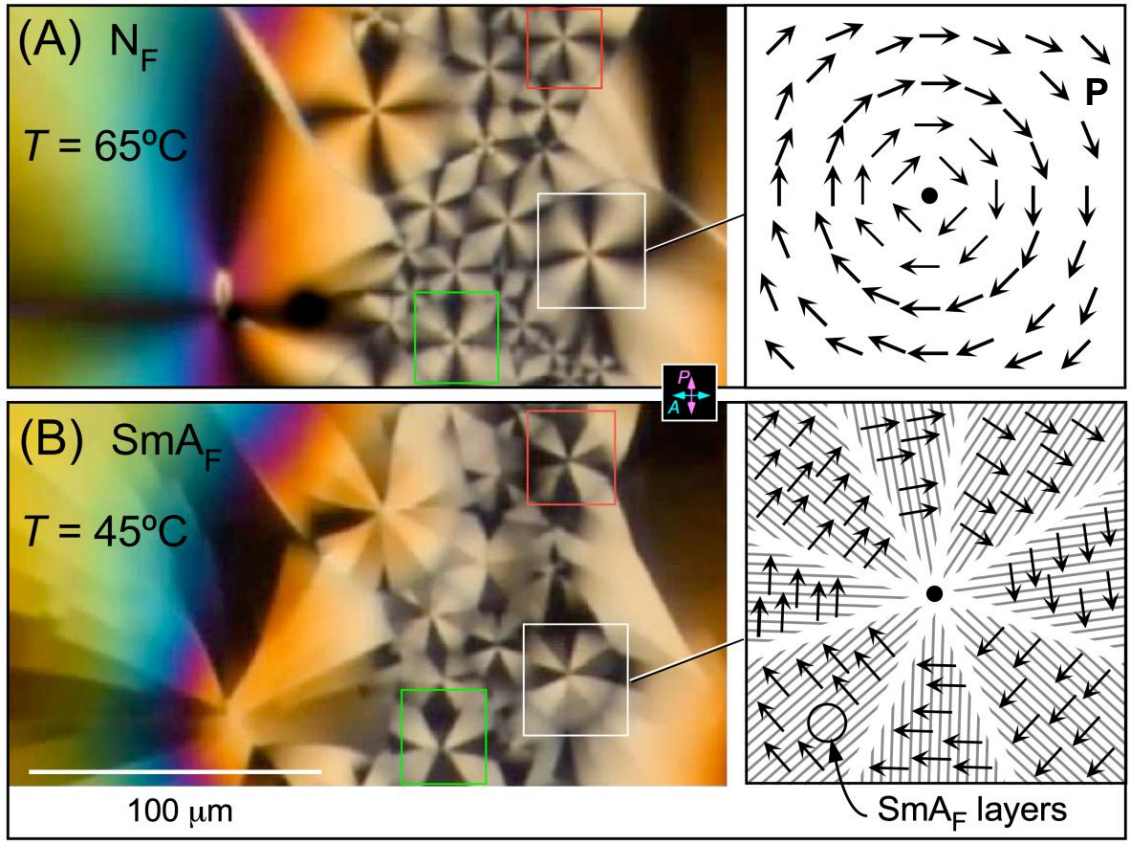

*Figure 3*. Polarized light microscope images of the starting and ending states of the director/polarization field around topological point defects, obtained on cooling a freely suspended 2N/DIO film from the N$_F$ to the SmA$_F$ phase at -10°C/min. Three selected defects are highlighted by colored squares. The variation in birefringence color in this part of the film shows that the film is thickest on the left ($d$~800 nm) and thinnest on the right, where the defects are aggregated ($d$~100 nm). (A) The preferred director/polarization field around +1 defects in the N$_F$ phase exhibits pure bend, making $\nabla \cdot P=0$ and eliminating polarization charge to minimize the electrostatic energy. (B) On cooling to the SmA$_F$ phase, the smoothly bent smectic layers straighten out, forming blocks with uniformly oriented polarization separated by grain boundaries. This arrangement minimizes the energy densities associated with both smectic layer bend elasticity, and space charge arising from polarization splay.

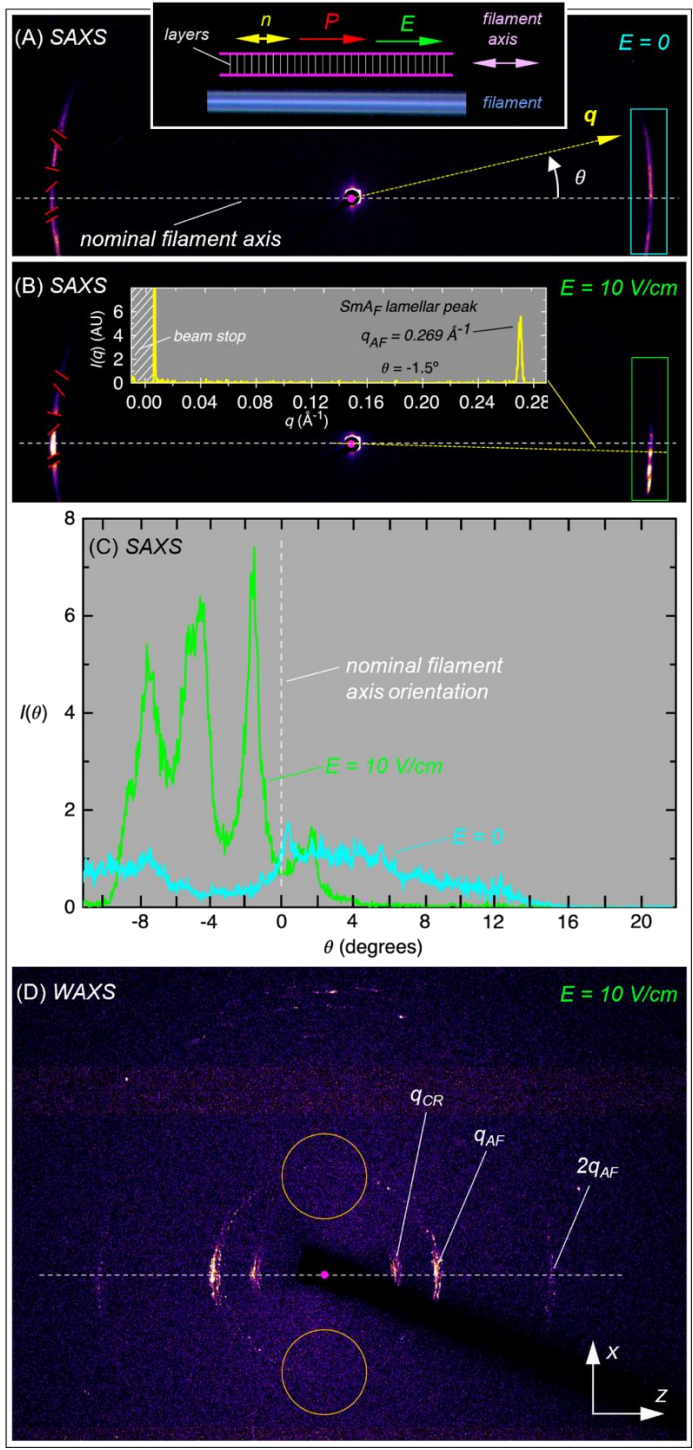

*Figure 4*. X-ray diffraction from freely suspended 2N/DIO filaments in the SmA$_F$ phase. SAXS images were obtained (*A*) in the absence of applied field and (*B*) with a weak electric field applied along the filament axis. The beam center is indicated by a magenta dot. The inset in *(A)* shows a freshly drawn filament 1 mm in diameter viewed in reflected light and a schematic model of the filament structure. The Bragg peaks at $q$ = 0.269 Å$^{-1}$, indicating a layer spacing $d_{AF}$ = 23.4 Å and shown in the line scan in *(B)*, are consistent with scattering from smectic layers oriented generally normal to the filament axis. Gaps in the diffraction arcs at left (demarcated with red lines) are artefacts due to wires partially occluding the scattered light in the evacuated beam path. (*C*) Angular distributions of intensity along the right-side SAXS scattering arc, $I(\theta)$, indicative of a mosaic distribution of layer-normal directions in the illuminated part of the filament. In the absence of applied field, the scattering is broad and azimuthally diffuse (cyan curve). Application of a field along the filament axis narrows the distribution and shows several well-defined peaks corresponding to oriented smectic domains. (*D*) WAXS of the same filament reveals both first- and second-harmonic Bragg scattering from the SmA$_F$ phase,

as well as first-order scattering from the crystal phase, which grows in slowly near room temperature. Significantly, there is no evidence of scattering at wavevector $q_{AF}$ along the $\boldsymbol{q}_x$ direction (inside the orange circles), indicating that there is no smectic layering on or parallel to the filament/air interfaces.

*Figure 5.* Freely suspended and free-standing 2N/DIO filaments in the SmA$_F$ phase viewed in reflection under crossed polarizers. Filaments are generally horizontal in the lab, with gravity (*g*) normal to the image plane, as indicated in (*A*). Filaments accommodate thickness changes along their length with alternating cylindrical and conical segments. The resultant variations in filament surface orientation and overall bending in gravity result in differences in reflectivity along the filament length. (*A,B*) show that the film reflectivity is depolarized, resulting from the internal birefringence, with optical axes along x,z. (*C–E*) Time sequence over ~60 sec in which a thin LC bridge tethering the right end of the filament

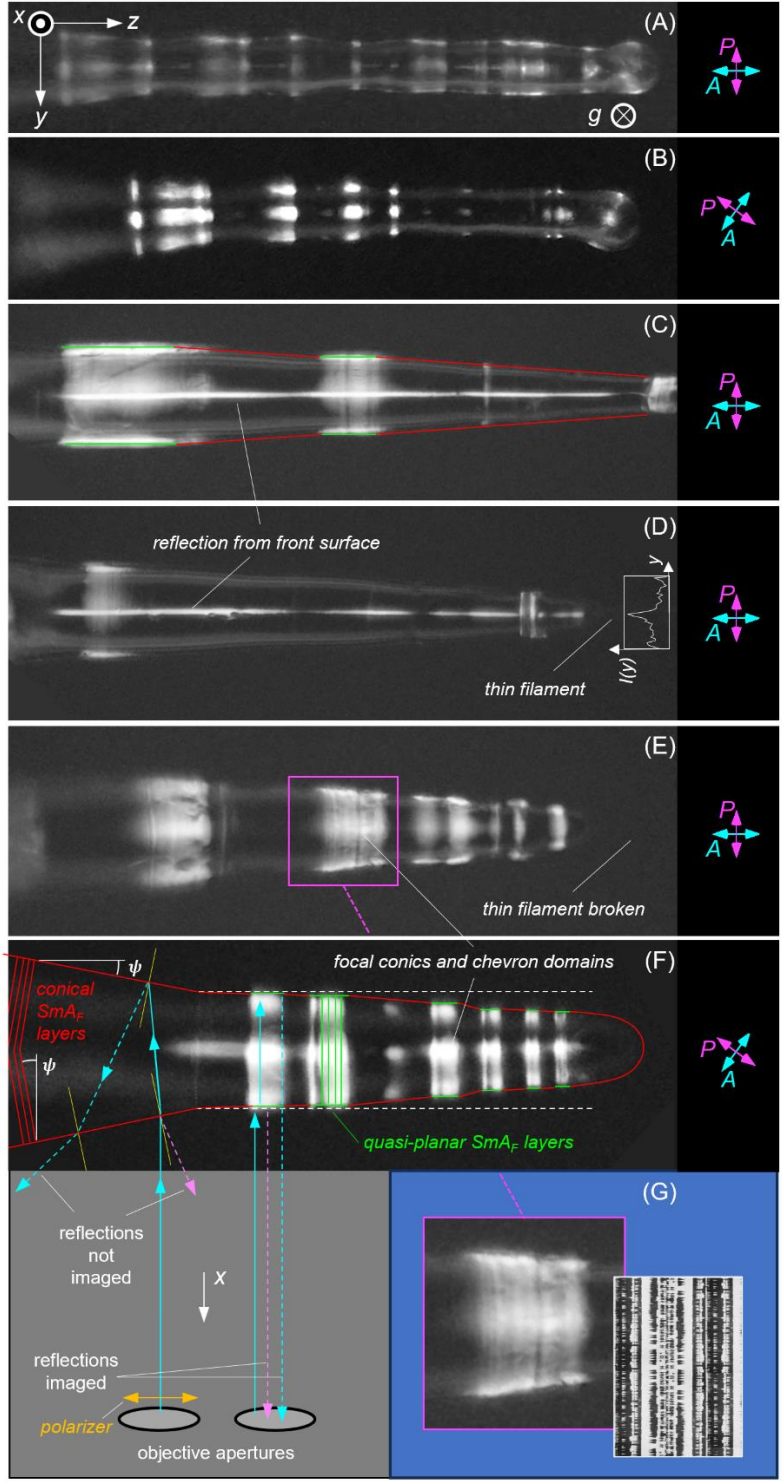

pinches off and breaks. The filament then bends down slightly under gravity, behind the image plane, but does not retract, illustrating the extraordinary mechanical stability of SmA$_F$ filaments. The thin bridge cannot be resolved here but the inset in (*D*) shows an intensity scan across its width. (*F*) Schematic illustration of the layer geometry and reflection optics of a filament with cylindrical (red) and conical (green) sections, showing the origin of the bright and dark bands in the filament. (*G*) Magnified view of the highly defected smectic texture of the conical filament region outlined in (E). The smaller image shows zig-zag layering defects in a chiral smectic C cell viewed in polarized transmission.

*Supplemental Information*
*Freely-Suspended Nematic Films and Free-Standing Smectic Filaments*
*in the Ferroelectric Nematic Realm*

Keith G. Hedlund, Vikina Martinez, Xi Chen, Cheol S. Park,

Joseph E. Maclennan, Matthew A. Glaser, and Noel A. Clark

*Department of Physics, University of Colorado, Boulder, Colorado, 80309, USA*

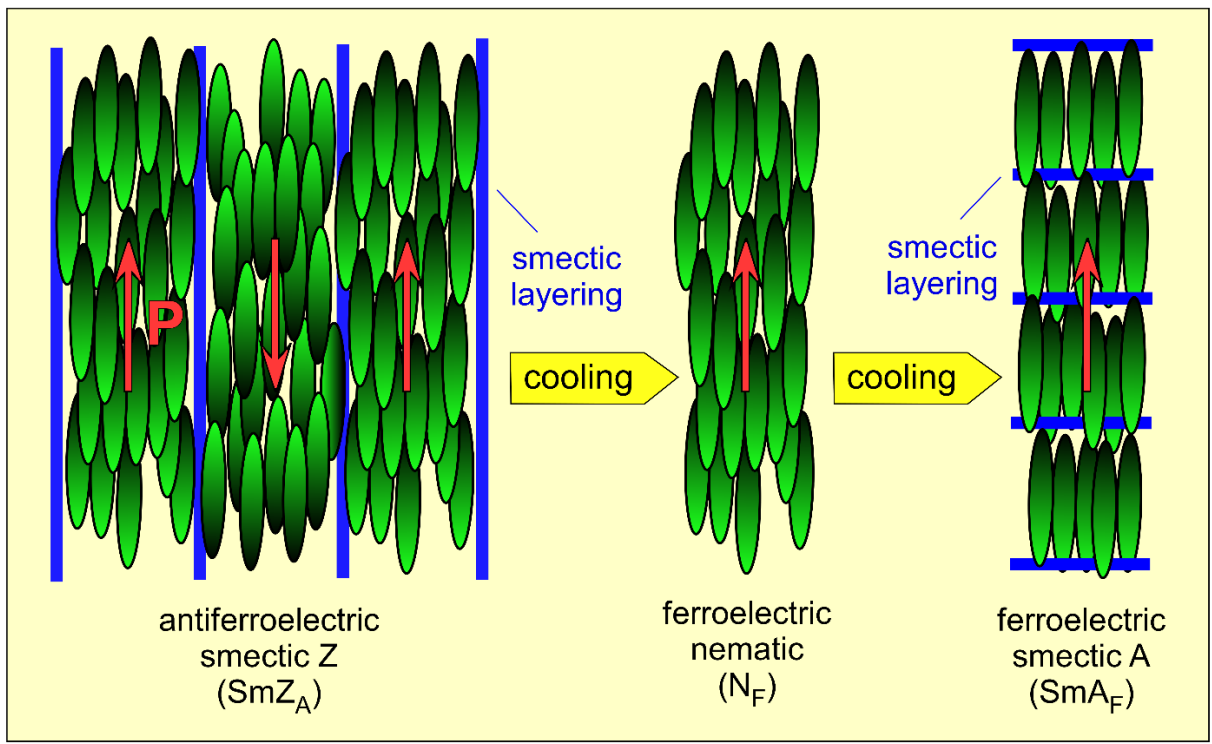

**Figure S1**. Three polar phases of the ferroelectric nematic realm. The SmZ$_A$ is an antiferroelectric, lamellar phase with the polarization **P** parallel to the layers and alternating in direction from layer to layer. The N$_F$ is the ferroelectric nematic phase, with the polarization uniform along the director. The SmA$_F$ is a ferroelectric, lamellar phase with the polarization also along the director.

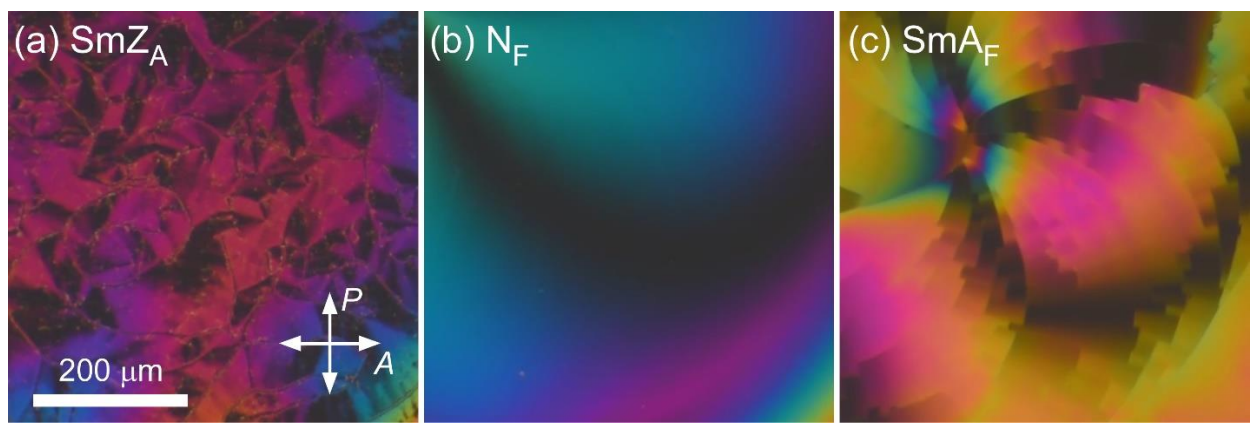

**Figure S2**. Typical textures of a freely suspended film of a 50:50 wt% mixture of AUUQU2N (2N) and DIO in the (a) SmZ$_A$ (85°C), (b) N$_F$ (68°C), and (c) SmA$_F$ (45°C) phases, viewed in transmission through crossed polarizers.

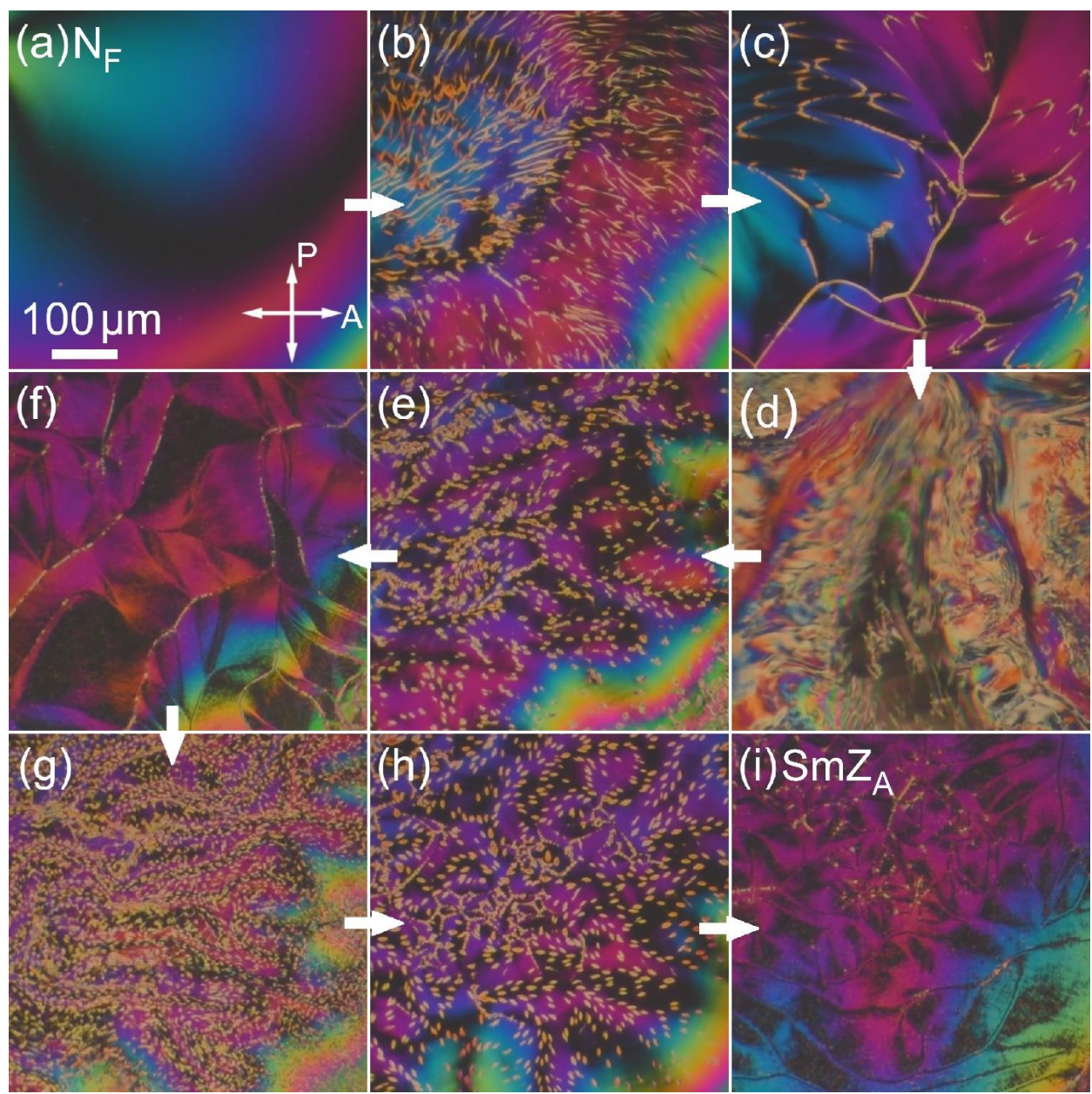

**Figure S3.** Transient texture observed in a slowly heated freely suspended film of a 50:50 wt% mixture of AUUQU2N (2N) and DIO at the transition from the $N_F$ to the $SmZ_A$ phase, at around 87°C. Small fluctuations in the film temperature in the vicinity of the transition result in the rapid, repeated formation and disappearance of the small, orange "rice grain" features that are characteristic of this first-order transition. Similar behavior can be seen both on heating, as in this sequence, and on cooling.

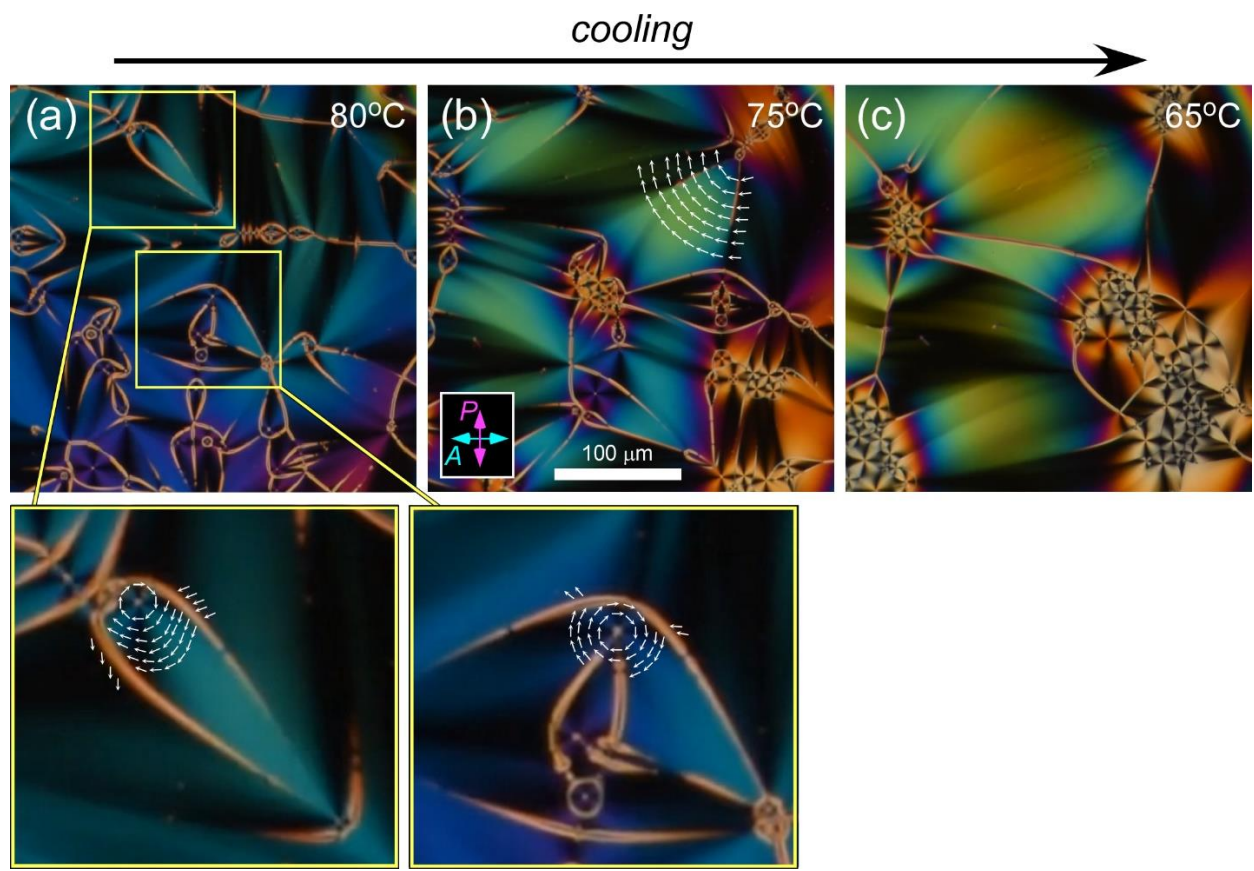

**Figure S4.** Point and line defects on a freely suspended smectic film of a 50:50 wt% mixture of AUUQU2N (2N) and DIO cooled at -10°C per minute in the N$_F$ phase. The polarization fields for two of the parabolic defects in (a) are sketched below the main figure. On cooling, the point defects assemble into clusters and the film in their vicinity becomes significantly thinner.

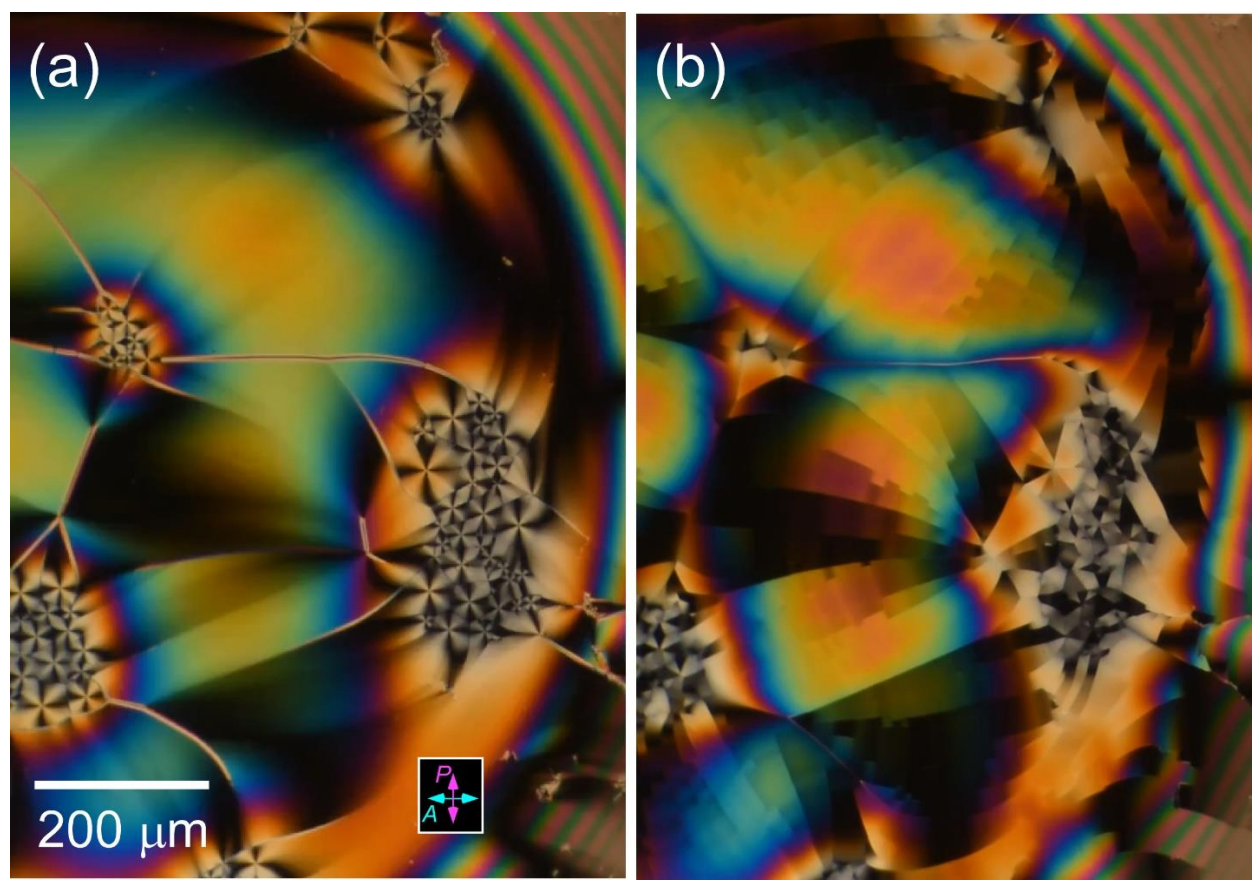

**Figure S5**. Formation of grain boundaries and evolution of a cluster of point defects in adjacent regions of a freely suspended smectic film of a 50:50 wt% mixture of AUUQU2N (2N) and DIO cooled at -10°C per minute from (a) the $N_F$ phase (65°C) to (b) the $SmA_F$ phase (45°C).

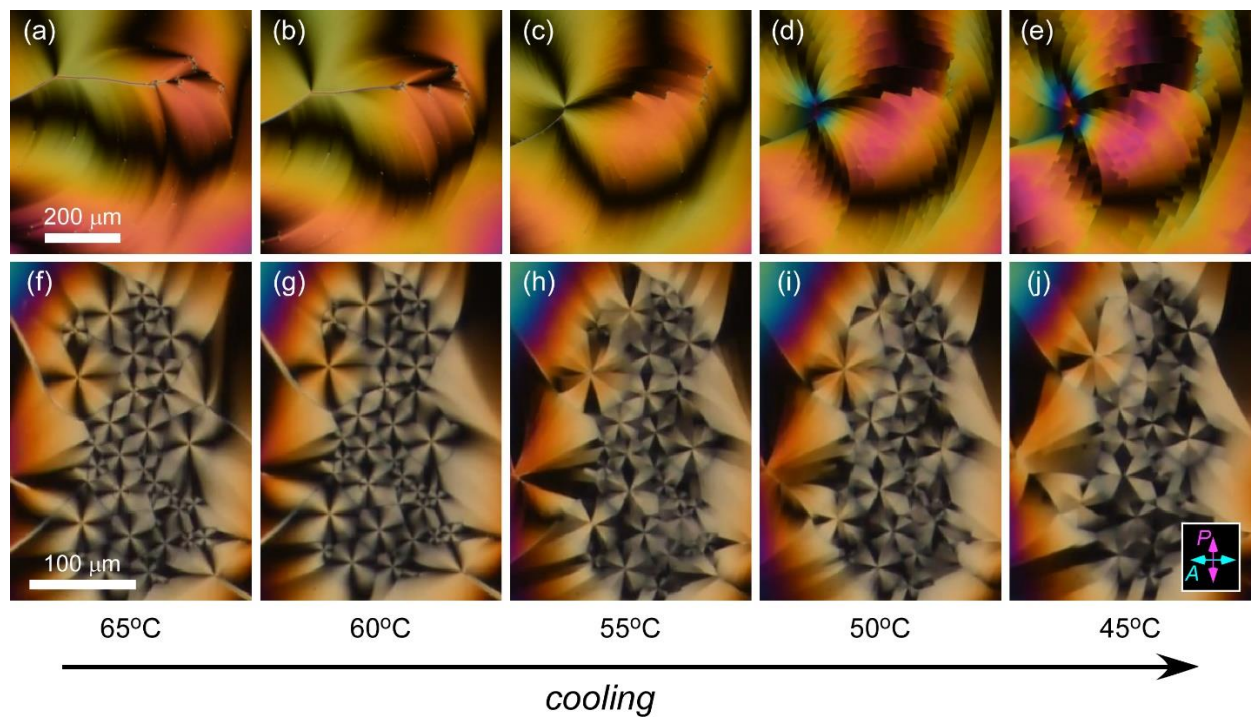

**Figure S6**. Detailed view of the formation of grain boundaries (a–e) and evolution of a cluster of point defects (f–j) in adjacent regions of the freely suspended smectic film shown in Fig. S3, cooled at -10°C per minute from the $N_F$ phase (a,f) to the $SmA_F$ phase (e,j).

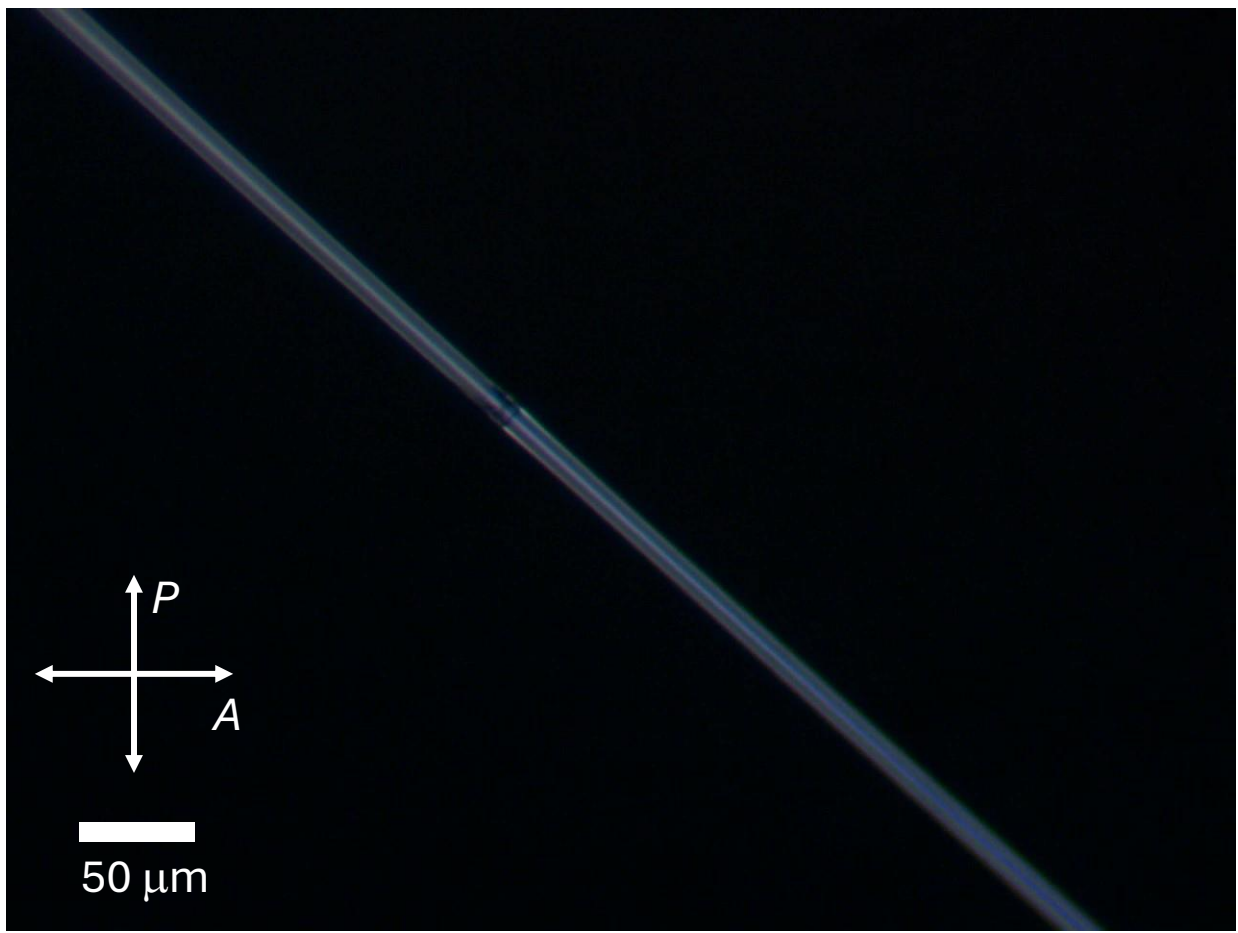

**Figure S7**. Typical section of a long, thin SmA$_F$ filament of a 50:50 wt% mixture of AUUQU2N (2N) and DIO at 35°C, viewed in reflected light at 20x magnification. This filament is approximately 4 mm in length and about 10 microns in diameter. The bright centerline is a reflection from the top surface of the filament. The narrow, dark band near the middle of the image corresponds to a conical region mediating a small change in filament diameter.

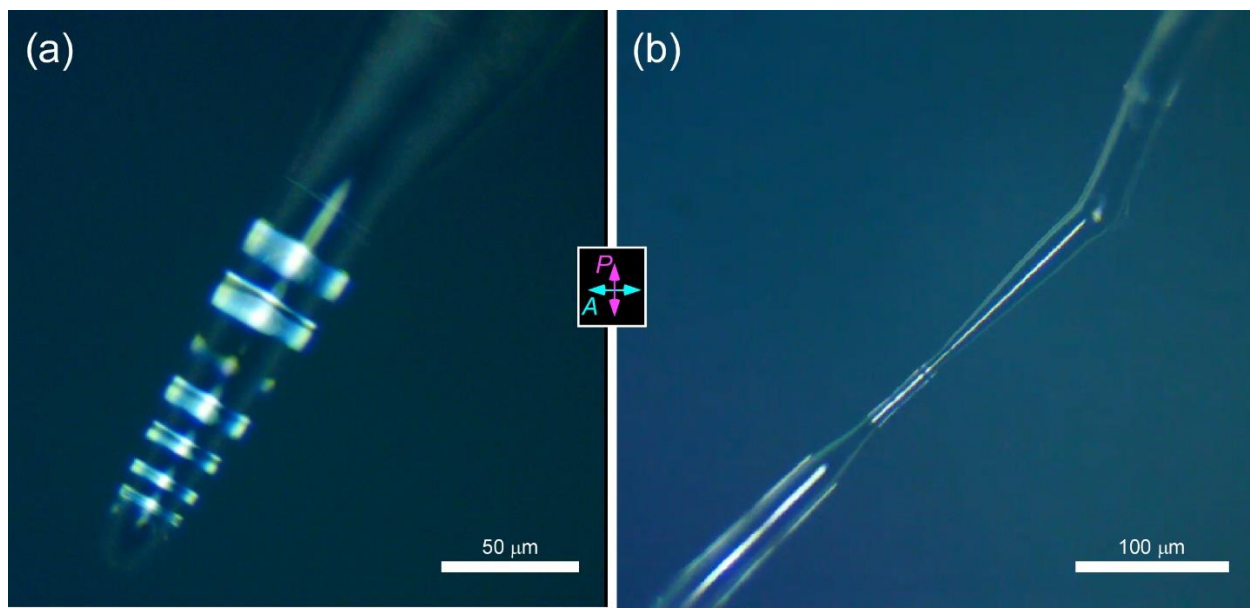

**Figure S8**. SmA$_F$ filaments of a 50:50 wt% mixture of AUUQU7N (2N) and DIO at 35°C, viewed in polarized reflected light at 20x magnification. The filaments show the typical banded textures and variations in reflectivity associated with changes in diameter and filament orientation. The upper part of the filament in (b) appears dark because it is bent downward, out of the image plane.

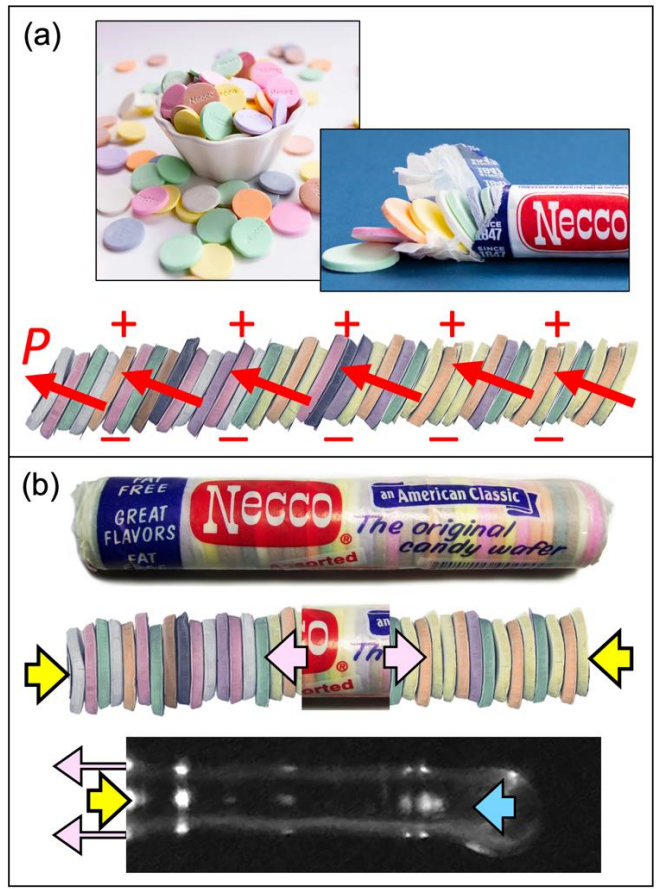

**Figure S9**. New England Confectionery Company candy wafers. (a) These disc-like, edible wafers are non-adhesive when dry but are frictional like chalk, making them also useful as building toys or as stacked tokens for games. (b) The wafers are arranged in single, columnar stacks in a gossamer-like paper tube, as shown. This packaging is robust, in spite of the lightweight nature of the tube. During production, the wafers are slipped into the tube and then compressed slightly on the ends (yellow forces). The tube is then crimped shut on the ends, in the process of which it is stretched by a small amount. This results in a dilative force (pink force, equal in magnitude to the yellow force) transmitted along the tube from end to end, which in turn puts the stack permanently under weak axial compression. This force suppresses the tilt of the wafers seen in (a), effectively minimizing the length of the stack (comparing the images in (a) and (b)). The actinic (axial) pink force stretches the cylindrical tube parallel to its axis, tending to keep its paper wall taut eveywhere and parallel to the axis, generating stresses that suppress relative displacement of the wafers in directions normal to the axis and keep them lined up next to each another. In an LC filament, shown in the final graphic, the equivalent of the dilative force is provided by surface tension (pink arrows), which also keeps the cross-section of the filament circular. The blue arrow represents Laplace pressure due to the surface tension around the bulbous free end of the filament.

9